\documentclass[electronic]{vgtc}  
\usepackage[utf8]{inputenc}
\usepackage{times}

\usepackage{amsmath}
\usepackage{xspace}
\usepackage{subcaption}
\usepackage{graphicx} 
\usepackage{multirow}
\usepackage{makecell}
\usepackage[normalem]{ulem}
\usepackage{wrapfig}
\usepackage{rotating}
\usepackage{soul}
\usepackage{threeparttable}
\usepackage{textgreek}
\usepackage{tabularx}
\usepackage[dvipsnames,table]{xcolor}
\usepackage[linesnumbered,ruled,vlined, noend]{algorithm2e}
\usepackage{booktabs}
\usepackage{siunitx}
\usepackage{enumitem}
\PassOptionsToPackage{
  colorlinks=true,
  linkcolor=black,
  citecolor=black,
  urlcolor=blue,
  hidelinks=false
}{hyperref}
\usepackage{wasysym}
\usepackage{amssymb}
\usepackage{etoolbox}

\usepackage{fontawesome5}

\vgtcinsertpkg

\usepackage{array}
\newcolumntype{Y}{>{\centering\arraybackslash}X} 
\newcolumntype{P}[1]{>{\raggedright\arraybackslash}p{#1}} 

\DeclareSIUnit{\foot}{ft}

\vgtccategory{Research}

\newcommand{\sysname}{\textbf{MURMR}\xspace}

\AtBeginDocument{
  \abovedisplayskip=2pt
  \belowdisplayskip=2pt
  \abovedisplayshortskip=2pt
  \belowdisplayshortskip=2pt
}
\everydisplay{\abovedisplayskip=2pt \belowdisplayskip=2pt}

\usepackage{caption}
\title{MURMR: A Multimodal Sensing Pipeline for Automated Group Behavior Analysis in Mixed Reality}

\author{
Diana Romero${}^{*}$ \\
dgromer1@uci.edu \\
University of California, Irvine\\
\and
Yasra Chandio\thanks{Diana Romero and Yasra Chandio contributed equally to this work as co-first authors.} \\
ychandio@umass.edu \\
University of Massachusetts Amherst\\
\and
Fatima M. Anwar \\
fanwar@umass.edu \\
University of Massachusetts Amherst\\
\and
Salma Elmalaki \\
salma.elmalaki@uci.edu \\
University of California, Irvine\\
}

\onlineid{1358}

\vspace{-.2cm}
\teaser{
  \centering
  \includegraphics[width=1\linewidth, trim={0.4cm 12.9cm 9.9cm 2cm},clip]{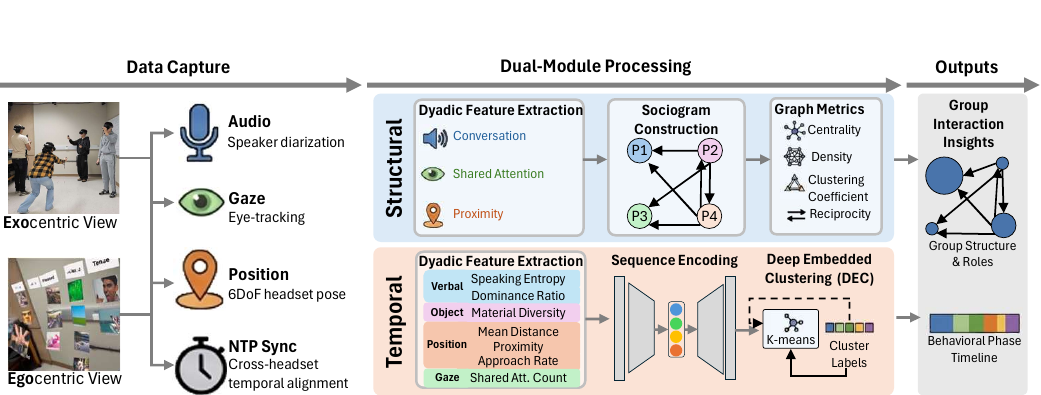}
  
  \caption{\sysname pipeline for sensing and analyzing collaborative group behavior in MR, starts by synchronizing multimodal sensor data to maintain consistency, then operates at two complementary modules. At a macro-level, session-long sensor data is aggregated to build sociograms that reveal overall group structure. At a micro-level, a temporal analysis of short interaction windows classifies fine-grained behavioral patterns. 
  } 
  \label{fig:sys-overview}
}

\abstract{

When teams coordinate in immersive environments, collaboration breakdowns can go undetected without automated analysis, directly affecting task performance. Yet existing methods rely on external observation and manual annotation, offering no annotation-free method for analyzing temporal collaboration dynamics from headset-native data. We introduce \sysname, a passive sensing pipeline that captures and analyzes multimodal interaction data from commodity MR headsets without external instrumentation. Two complementary modules address different levels of analysis: a structural module that generates automated multimodal sociograms and network metrics at both session and intra-session granularities, and a temporal module that applies unsupervised deep clustering to identify moment-to-moment dyadic behavioral phases without predefined taxonomies.

An exploratory deployment with 48 participants in a co-located object-sorting task reveals that intra-session structural analysis captures significant within-session variability lost in session-level aggregation, with gaze, audio, and position contributing non-redundantly. The temporal module identifies five behavioral phases with 83\% correspondence to video observations. Cross-tabulation shows that behavioral transitions consistently occur within structurally stable states, demonstrating that the two modules capture complementary dynamics. These results establish that passive headset sensing provides meaningful signal for automated, multi-level collaboration analysis in immersive environments.

}


\begin{document}

\maketitle

\section{Introduction}\label{sec:intro}
Collaborative Mixed Reality (MR) applications are transforming fields such as surgical training~\cite{gerup_augmented_2020}, shared 3D architectural walkthroughs~\cite{du2016communication-architecture-walkthroughs}, remote industrial equipment maintenance guidance~\cite{gonzalez-franco_immersive_2017}, and co-creative product prototyping~\cite{cascini2020exploring-prototype-creative}, all of which rely on seamless coordination among multiple users in real time. When collaboration breaks down in these settings, whether through miscommunication during a surgical procedure or a lapse in coordination during remote maintenance, the consequences directly compromise safety, task outcomes, and team performance~\cite{paulus2000groups,stogdill1972group}. Yet such breakdowns often go undetected without trained specialists or extensive post-hoc video review, neither of which scale to routine use. Commodity MR headsets, however, already capture gaze direction, spatial position, and audio as a byproduct of normal operation. Making this an underexploited data source for automated collaboration analysis.

Prior work in computer-supported cooperative work (CSCW) and human-computer interaction (HCI) has pursued two main avenues toward understanding collaboration in immersive environments. The first builds tools that help specialists observe and analyze group behavior, for example through replaying recorded sessions and manually coding interaction patterns~\cite{yangImmersiveCollaborativeSensemaking2022}. While effective, these approaches depend on expert availability and post-hoc annotation. The second avenue pursues automated in-situ analysis, including temporal methods such as process mining and sequence analysis, but these typically rely on external infrastructure such as motion capture systems~\cite{zhang2022real-3dvide}, wearable sensors like sociometric badges~\cite{kimSociometricBadgesUsing2012, zhang_teamsense_2018}, or predefined behavioral categories~\cite{lechappe2025categorize} validated only under scripted task conditions. No existing approach passively derives both the structural organization and the temporal behavioral dynamics of group collaboration from commodity MR headset sensors alone, without relying on external hardware or annotation.

We investigate this gap through three research questions:

\begin{itemize}[leftmargin=0.5cm, topsep=-0.1cm, noitemsep]
    \item \textbf{RQ1}: How do group collaboration patterns vary across temporal granularities, and what dynamics does session-level aggregation obscure?
    \item \textbf{RQ2}: What distinct behavioral phases characterize moment-to-moment dyadic interaction in co-located MR collaboration?
    \item \textbf{RQ3}: To what extent do structural and temporal analyses capture complementary aspects of collaboration?
\end{itemize}

We approach these questions through \sysname{} (\underline{M}ultimodal \underline{U}nsupervised \underline{R}elational \underline{MR}), a passive sensing pipeline that combines structural and temporal analysis of multimodal data from commodity MR headsets. Our contributions are:

\begin{enumerate}[leftmargin=0.5cm, topsep=-0.1cm, noitemsep]
    \item The first \emph{passive multimodal sensing pipeline} that derives group collaboration analytics exclusively from commodity MR headset sensors, without external hardware or annotation.
    \item A \emph{structural analysis module} that translates multimodal sensor streams into weighted sociograms at both session and intra-session granularities, revealing within-session variability in group organization that session-level aggregation obscures.
    \item An \emph{unsupervised temporal clustering module} that identifies distinct behavioral phases from multimodal dyadic interaction streams without predefined categories, surfacing moment-to-moment dynamics inaccessible to session-level methods.
    \item An \emph{exploratory deployment} with 48 participants demonstrating that structural and temporal analyses capture complementary dynamics: behavioral transitions occur within structurally stable states, indicating that both levels of analysis are necessary for characterizing collaboration.
\end{enumerate}

\newcolumntype{P}[1]{>{\RaggedRight\arraybackslash}p{#1}}
\renewcommand\theadalign{bc}

\begin{table*}[!t]
  \centering
  \caption{Comparative analysis of related works across nine dimensions of collaborative MR analytics.
    Columns are organized around three capability tiers: behavioral inference (collaboration metrics,
    automated metrics, unsupervised inference, temporal analysis), deployment model (no external
    hardware, annotation-free), and scope (group-level, subgroup detection, multi-user$^*$).
    Symbols: \CIRCLE~=~present, \LEFTCIRCLE~=~partially present or limited, \Circle~=~absent.}
  \label{tab:comparison}
  \scriptsize
\begin{tabularx}{\textwidth}{|
  Y|
  *{1}{>{\centering\arraybackslash}p{0.78cm}|}
  *{1}{>{\centering\arraybackslash}p{0.78cm}|}
  *{1}{>{\centering\arraybackslash}p{1.6cm}|}
  *{1}{>{\centering\arraybackslash}p{1.6cm}|}
  *{1}{>{\centering\arraybackslash}p{1.0cm}|}
  *{1}{>{\centering\arraybackslash}p{1.1cm}|}
  *{1}{>{\centering\arraybackslash}p{0.78cm}|}
  *{1}{>{\centering\arraybackslash}p{1.2cm}|}
  *{1}{>{\centering\arraybackslash}p{0.78cm}|}
}
\hline
\thead{Work} &
\rotatebox{0}{\thead{Collab.\\metrics}} &
\rotatebox{0}{\thead{Auto.\\metrics}} &
\rotatebox{0}{\thead{Unsupervised\\inference}} &
\rotatebox{0}{\thead{Temporal\\analysis}} &
\rotatebox{0}{\thead{No ext.\\hardware}} &
\rotatebox{0}{\thead{Annotation\\-free}} &
\rotatebox{0}{\thead{Group-\\level}} &
\rotatebox{0}{\thead{Subgroup\\detection}} &
\rotatebox{0}{\thead{Multi-\\user$^*$}} \\ \hline

\rowcolor{green!30}
\sysname~(ours)                                                           & \CIRCLE      & \CIRCLE      & \CIRCLE      & \CIRCLE      & \LEFTCIRCLE & \CIRCLE      & \CIRCLE      & \CIRCLE      & \CIRCLE      \\ \hline \hline

\multicolumn{10}{|l|}{\textit{Automated collaboration analytics (\S\ref{sec:analytics})}} \\ \hline
Yang~(2022)~\cite{yangImmersiveCollaborativeSensemaking2022}              & \CIRCLE      & \Circle      & \Circle      & \Circle      & \CIRCLE     & \Circle      & \CIRCLE      & \Circle      & \LEFTCIRCLE  \\ \hline
Echeverria~(2019)~\cite{echeverria2019towards}                            & \CIRCLE      & \LEFTCIRCLE  & \Circle      & \LEFTCIRCLE  & \Circle     & \Circle      & \CIRCLE      & \Circle      & \CIRCLE      \\ \hline
L\'{e}chapp\'{e}~(2025)~\cite{lechappe2025categorize}                    & \CIRCLE      & \LEFTCIRCLE  & \Circle      & \LEFTCIRCLE  & \LEFTCIRCLE & \LEFTCIRCLE  & \CIRCLE      & \Circle      & \CIRCLE      \\ \hline
Nguyen~(2013)~\cite{nguyen2013extraction}                                        & \LEFTCIRCLE  & \CIRCLE      & \CIRCLE      & \Circle      & \Circle     & \CIRCLE      & \CIRCLE      & \LEFTCIRCLE  & \CIRCLE      \\ \hline
Huang~(2019)~\cite{huang2019identifying}                                & \CIRCLE      & \CIRCLE      & \CIRCLE      & \LEFTCIRCLE  & \Circle     & \LEFTCIRCLE  & \CIRCLE      & \Circle      & \CIRCLE      \\ \hline
Ma~(2021)~\cite{ma2021unsupervised}                                       & \Circle      & \CIRCLE      & \CIRCLE      & \LEFTCIRCLE  & \LEFTCIRCLE & \CIRCLE      & \Circle      & \Circle      & \Circle      \\ \hline \hline

\multicolumn{10}{|l|}{\textit{Sensing, replay, and infrastructure (\S\ref{sec:sensing})}} \\ \hline
\textsc{ISA}~(2024)~\cite{lammert2024immersive}                           & \LEFTCIRCLE  & \LEFTCIRCLE  & \Circle      & \LEFTCIRCLE  & \Circle     & \Circle      & \CIRCLE      & \LEFTCIRCLE  & \CIRCLE      \\ \hline
\textsc{ReLive}~(2022)~\cite{hubenschmid2022relive}                       & \Circle      & \Circle      & \Circle      & \Circle      & \LEFTCIRCLE & \LEFTCIRCLE  & \LEFTCIRCLE  & \Circle      & \CIRCLE      \\ \hline
\textsc{MIRIA}~(2021)~\cite{buschel2021miria}                             & \Circle      & \Circle      & \Circle      & \Circle      & \CIRCLE     & \Circle      & \LEFTCIRCLE  & \Circle      & \CIRCLE      \\ \hline
\textsc{AutoVis}~(2023)~\cite{jansen2023autovis}                          & \Circle      & \Circle      & \Circle      & \Circle      & \Circle     & \LEFTCIRCLE  & \LEFTCIRCLE  & \Circle      & \LEFTCIRCLE  \\ \hline
\textsc{Tesseract}~(2023)~\cite{mahadevan2023tesseract}                   & \Circle      & \Circle      & \Circle      & \Circle      & \Circle     & \Circle      & \Circle      & \Circle      & \Circle      \\ \hline
\textsc{PLUME}~(2024)~\cite{javerliat2024plume}                           & \Circle      & \Circle      & \Circle      & \Circle      & \LEFTCIRCLE & \LEFTCIRCLE  & \Circle      & \Circle      & \Circle      \\ \hline
\textsc{PSI}~(2021)~\cite{bohus2021platform}                              & \Circle      & \Circle      & \Circle      & \Circle      & \Circle     & \Circle      & \Circle      & \Circle      & \CIRCLE      \\ \hline
\textsc{MRAT}~(2020)~\cite{nebeling2020mrat}                              & \Circle      & \Circle      & \Circle      & \Circle      & \LEFTCIRCLE & \Circle      & \LEFTCIRCLE  & \Circle      & \CIRCLE      \\ \hline

\end{tabularx}
\vspace{1pt}
\raggedright\scriptsize{$^*$\textit{Multi-user} denotes native support for analyzing inter-user collaboration dynamics, not merely recording multiple simultaneous users.}
\end{table*}
\section{Related Work}
We organize related work around three threads: behavioral sensing systems in MR (\S\ref{sec:sensing}), analytical methods for collaboration pattern extraction (\S~\ref{sec:analytics}), and a comparative positioning of \sysname{} against existing approaches (\S~\ref{sec:position}).

\vspace{-0.3em}
\subsection{Group Behavior Sensing in VR/MR}\label{sec:sensing}

Research on sensing group behavior originated in physical settings, where wearable and mobile sensors, including sociometric badges, body-worn accelerometers, and microphones, were used to capture face-to-face interaction patterns and infer team cohesion~\cite{zhang_teamsense_2018, olguinCapturingIndividualGroup2009, kimSociometricBadgesUsing2012, pentland2012new}. This body of work established that commodity sensors can surface rich group dynamics, but the reliance on dedicated external hardware limits scalability and deployability outside controlled lab environments.

Within immersive environments, a growing ecosystem of tools addresses different stages of the behavioral data lifecycle. In-situ and immersive replay tools such as MIRIA~\cite{buschel2021miria}, ReLive~\cite{hubenschmid2022relive}, and ISA~\cite{lammert2024immersive} enable researchers to re-enter recorded environments for observation and behavioral coding, with domain-specific extensions such as AutoVis~\cite{jansen2023autovis} and Tesseract~\cite{mahadevan2023tesseract} demonstrating the value of immersive data review in automotive and design contexts. Infrastructure systems such as PLUME~\cite{javerliat2024plume}, PSI~\cite{bohus2021platform}, and MRAT~\cite{nebeling2020mrat} provide capture, synchronization, and standardized formats for XR behavioral data, but do not infer group-level social dynamics from the data they collect. Simulation-based approaches such as MoCoMR~\cite{romero2025mocomr} complement this ecosystem by generating synthetic collaborative MR data from learned individual behavior models, though they target data augmentation rather than in-situ behavioral inference.

Across both lines of work, collaboration pattern extraction remains manual and analyst-driven. \sysname{} is complementary to these tools: it operates on the same headset-native sensor streams they capture but shifts the focus from replay and manual coding to automated structural and temporal analysis of group behavior.

\vspace{-0.3em}
\subsection{Automated Collaboration Analytics}\label{sec:analytics}
A longstanding tradition in small group research uses Social Network Analysis (SNA) and sociometry, translating observed behavioral signals into sociograms that visualize relationships, roles, and subgroup structures~\cite{moreno1941foundations,wasserman1994social-faust-1994}. This approach has been applied across domains including nursing teams~\cite{drahota2008sociogram-as-an-analytical-tool}, classrooms~\cite{martinezCombiningQualitativeEvaluation2003}, and VR settings where interaction and gaze patterns reflect meaningful differences in collaborative behavior~\cite{yangImmersiveCollaborativeSensemaking2022,bai2020user}. However, most applications construct sociograms as static, session-level aggregates.

Temporal analysis methods in CSCW and learning analytics capture how collaboration evolves over time rather than summarizing it in aggregate. Methods such as process mining~\cite{van2012process}, sequence analysis~\cite{malmberg2017capturing}, and epistemic network analysis~\cite{zhao2024epistemic} extract dynamic patterns from interaction logs, but require predefined state labels or manually coded categories as input. Systems that automate collaboration assessment face analogous constraints: Echeverria et al.~\cite{echeverria2019towards} fuse positioning, audio, and physiological data into visual proxies organized around predefined theoretical dimensions, while L\'echapp\'e et al.~\cite{lechappe2025categorize} pursue real-time detection of collaboration profiles from multimodal VR signals but validate against scripted dyadic scenarios with known behavioral categories. Across both temporal and assessment-oriented approaches, applicability is limited to settings where behavioral categories are established in advance and task conditions can be controlled.

Unsupervised representation learning offers an alternative by discovering latent behavioral patterns without pre-specified state definitions. Ma et al.~\cite{ma2021unsupervised} proposed a CNN-BiLSTM autoencoder combined with K-means clustering to discover human activity categories from unlabeled wearable sensor streams, demonstrating that deep representations can recover semantically meaningful behavioral groupings without annotation. Nguyen et al.~\cite{nguyen2013extraction} applied hierarchical Dirichlet processes to sociometric badge data to infer latent interaction patterns, showing that social contexts can similarly be extracted unsupervised. In a collaborative desktop setting, Huang et al.~\cite{huang2019identifying} applied unsupervised clustering to multimodal sensor data, discovering collaborative states correlated with task performance, though using external sensors without structural network analysis. Across these works, unsupervised behavioral discovery remains largely disconnected from sociometric methods: studies that cluster interaction patterns do not typically construct sociograms, and studies that build sociograms do not typically discover temporal states from the data.

In summary, SNA provides structural characterization but typically at session-level granularity. Temporal and assessment methods capture dynamics but require predefined categories. Unsupervised learning discovers patterns but has not been integrated with sociometric analysis. No prior work combines structural and temporal analysis to discover multi-level collaboration patterns from passive MR headset data without predefined behavioral taxonomies.


\vspace{-0.3em}
\subsection{Positioning and Comparison}\label{sec:position}
We compare the systems discussed in \S\ref{sec:sensing} and \S\ref{sec:analytics} across nine capability dimensions spanning behavioral inference, deployment model, and scope (\autoref{tab:comparison}). Two caveats are reflected in the comparison: first, \sysname's sensing is headset-native but its analysis currently runs on an external server, which we mark as partial for the ``No ext.\ hardware'' dimension. Second, while the pipeline's modular design supports future real-time deployment, the current study evaluates offline processing only.

\section{System Design and Implementation}
\label{sec:system-design}

MURMR's architecture is driven by a single constraint: derive both the structural organization and temporal behavioral dynamics of group collaboration entirely from commodity headset sensors, without external hardware or annotation. The pipeline achieves this in three stages (\autoref{fig:sys-overview}). First, a passive sensing module synchronizes gaze, audio, and position data across headsets. The structural analysis module then converts these streams into weighted interaction graphs (automated sociograms) and derives network metrics that characterize group organization at both session and intra-session granularities. Finally, a temporal analysis module encodes short dyadic interaction segments into latent representations and clusters them to identify recurring behavioral phases without predefined categories. Together, the two analytical modules address complementary levels of collaboration that neither captures alone: stable relational structure and moment-to-moment behavioral variability.

\vspace{-0.3em}
\subsection{Design Rationale}\label{sec:design-rationale}
\noindent\textbf{Sensor scope.}
We restrict input to three modalities available on commodity MR headsets with built-in eye tracking, spatial tracking, and microphone capabilities: gaze direction, spatial position, and audio. This eliminates the need for external infrastructure and allows deployment in any setting where participants wear standard headsets. These three signals form a minimum viable sensor set: prior work has established that conversational turn-taking~\cite{pentland2012new}, spatial co-presence~\cite{hall1968proxemics}, and shared visual attention~\cite{wolfJointAttentionShared2016} each carry meaningful and complementary information about group coordination. Richer modalities such as hand tracking, body pose, or egocentric video could extend the pipeline in future work, but are not required for the structural and temporal analyses we target here. Cross-modal dependency analysis in \S\ref{sec:cross-modal} examines whether these three modalities contribute redundant or complementary information to the fused interaction representation.

\noindent\textbf{Output design.}
\sysname\ produces structured, machine-readable representations of group collaboration: sociogram edge weights, network metrics (centrality, density, reciprocity), cluster assignments, and temporal phase timelines. The present study evaluates the pipeline's analytical capabilities rather than its integration into specific end-user workflows, though the outputs are designed to support downstream applications such as real-time instructor dashboards, post-session collaborative debriefs, and adaptive task scaffolding in MR environments.

\vspace{-0.3em}
\subsection{Multimodal Passive Sensing Module}
\label{sec:passive-sensing}
\sysname is implemented as a lightweight Unity package that operates as a background service within the Unity runtime, interfacing with the headset's eye-tracking, audio, and spatial tracking APIs without modifying the host application's logic or rendering pipeline. During each MR session, the module captures three synchronized data streams from each headset. Integration requires only importing the package and initializing the sensing service at session start, with no scene modification required. To maintain temporal consistency across headsets, we perform clock alignment using the Network Time Protocol (NTP)~\cite{mills2002ntp}, achieving sub-\SI{100}{\milli\second} precision. This is sufficient because temporal overlap is accumulated across frames rather than requiring frame-exact coincidence. The finest-grained events in the pipeline, gaze convergence, operate at \SI{50}{\milli\second}--\SI{100}{\milli\second} timescales~\cite{yang2003short}.

Each sensor stream maps onto one of three dyadic interaction constructs used throughout the pipeline: audio produces directed \textit{conversation} edges encoding speaking-listening duration, gaze produces \textit{joint attention} edges capturing concurrent fixation on virtual objects, and position produces \textit{proximity} edges encoding physical co-presence. These constructs serve as the building blocks for both the structural sociograms (\S\ref{sec:structural}) and the temporal feature vectors (\S\ref{sec:temporalclustering}).

\noindent\textbf{{Audio.}} We capture speech via the headset's onboard microphone, recording at \SI{44.1}{\kilo\hertz} and processing with Pyannote~\cite{pyannote} for speaker diarization and voice activity detection. This produces timestamped speech segments with speaker identity for each participant. Segments shorter than \SI{0.5}{\second} are filtered, as prior work indicates this falls below the minimum duration at which listeners extract contextual meaning from speech~\cite{overath2015cortical} (formalized in \S\ref{sec:implementation_details}).

\noindent\textbf{{Gaze.}}
We capture gaze ray direction from each headset's built-in eye tracker, timestamped to the hardware clock. For each frame, gaze rays are cast against the bounding volumes of virtual objects in the scene, and a fixation on an object is registered when the ray intersects its hit box. Joint attention between two participants is detected when both users fixate on the same virtual object within overlapping time windows of at least \SI{13}{\milli\second}, a conservative lower bound adopted from vergence latency literature~\cite{yang2003short}. Overlapping gaze durations accumulate per dyad to form attention edge weights (formalized in \S\ref{sec:implementation_details}).

\noindent\textbf{{Position.}}
We record six-degree-of-freedom (6DoF) headset poses at \SI{1}{\hertz} intervals. All headsets are registered to a shared coordinate frame established at session start. In our implementation, this uses Photon Unity Networking, where one headset creates a spatial anchor at a known physical location and all other headsets align to it, though any networking stack that provides shared spatial anchoring could serve this role. The \SI{1}{\hertz} logging rate is consistent with the seconds-to-minutes timescale typical of co-presence events in collaborative tasks. Derived features such as approach rate are computed at this granularity, which limits sensitivity to rapid positional changes (\S\ref{sec:discussion}). Proximity interactions are recorded whenever pairwise headset distance falls within \SI{50}{\centi\meter}, corresponding to Hall's intimate distance zone~\cite{hall1968proxemics}, with accumulated co-presence duration serving as the edge weight (formalized in \S\ref{sec:implementation_details}).

\vspace{-0.3em}
\subsection{Structural Analysis Module}\label{sec:structural}

Dyadic interaction is inherently relational, existing between pairs of participants rather than within individuals, making weighted graphs a natural representation that preserves the strength and directionality of behavioral ties~\cite{moreno1941foundations,wasserman1994social}. In \sysname, we translate the synchronized sensor streams into a series of such graphs (automated sociograms), where each node is a \emph{participant} and the weight of each edge reflects the cumulative interaction strength across audio, gaze, and position modalities.

\vspace{-0.3em}
\subsubsection{Modality-Specific and Fused Sociogram Construction}\label{sec:sociogram_construction}

We construct a separate sociogram for each modality to isolate behaviors that could otherwise be conflated (for example, participants who remain physically close yet do not speak) and to allow per-modality analysis even when a sensor stream is temporarily unavailable. The three modality-specific sociograms (conversation, joint attention, and proximity) are then merged into a \emph{fused multimodal sociogram}.

\noindent\textbf{Analysis granularity.}
We construct sociograms at two granularities: session-level aggregations capturing overall interaction patterns, and sliding \SI{32}{\second} windows with \SI{16}{\second} stride for micro-level dynamic intra-session analysis. The \SI{32}{\second} window duration provides sufficient interaction events to generate stable edge weights while remaining sensitive to behavioral transitions that typically occur on sub-minute timescales in collaborative tasks (parameter selection detailed in \S\ref{sec:window_stride}).

\noindent\textbf{Edge definition.}
In each window, we assign edge weights based on the total \emph{duration} of interaction: the sum of spoken time for conversation, the overlap of gaze fixations for joint attention, and the accumulated intervals of co-presence for proximity (thresholds defined in \S\ref{sec:passive-sensing}). Conversation graphs are directed, reflecting the asymmetry between speaker and listener. Joint attention and proximity networks are undirected, as these forms of engagement are inherently mutual.

\noindent\textbf{Fused graph.}
After constructing three separate adjacency matrices, we z-score each and fuse them into a single multimodal network using PCA-derived weights, so that each channel contributes according to its shared variance across dyadic edges rather than equal or ad hoc contributions, while retaining the directed nature of conversational ties (formalized in \S\ref{sec:implementation_details}). To verify that this linear projection does not miss meaningful nonlinear structure, we compared against kernel PCA variants. Results in \S\ref{sec:cross-modal} confirm that the fused sociogram is stable across fusion methods.

\noindent\textbf{Cross-modal dependency analysis.}
To assess whether the three modalities carry redundant or complementary information, we compute conditional probabilities between modality-specific edges (e.g., $P(\text{joint attention} \mid \text{proximity})$) and conduct a leave-one-out ablation, reconstructing the fused sociogram from only two modalities and measuring rank correlation against the full three-modality fusion. This analysis provides empirical justification for the combined representation rather than assuming complementarity a priori. Results are reported in \S\ref{sec:cross-modal}.

\begin{table*}[!htbp]
\centering
\scriptsize
\caption{Interpretation of network metrics by interaction mode. \faMicrophone: conversation, \faEye: joint attention, \faMapPin: proximity.}
\vspace{-0.2cm}
\label{tab:network_metrics}
\begin{threeparttable}
\begin{tabularx}{\linewidth}{@{}| c| p{1.9cm}| X | X |@{}}
\hline
\textbf{Modality} & \textbf{Metric (Ref.)} & \textbf{High Value Interpretation} & \textbf{Low Value Interpretation} \\
\hline
\rowcolor[gray]{0.95}
\multicolumn{4}{@{}c}{\textbf{Centrality measures} identify potential leaders and information brokers in conversation, attention, and proximity networks.} \\
\hline
\faMicrophone, \faEye, \faMapPin & Eigenvector~\cite{freeman1979centrality} & Connected to other highly central participants& Linked mainly to peripheral participants \\
\hline
\rowcolor[gray]{.95}
\multicolumn{4}{@{}c}{\textbf{Cohesion measures} quantify bonding and tightness in proximity and attention networks.} \\
\hline
\faEye, \faMapPin & Clustering Coef. ~\cite{watts1998collective} & Neighbors are densely interconnected, reflecting tight local subgroup cohesion& Sparse neighbor connections, reflecting weak local cohesion \\
\hline
\faMicrophone, \faEye, \faMapPin & Density~\cite{scott2011social} & Well-connected network with active group engagement& Fragmented or minimally interacting group \\
\hline
\rowcolor[gray]{0.95}
\multicolumn{4}{@{}c}{\textbf{Connectivity measures} assess the balance of two-way exchanges in the conversation network.} \\
\hline
\faMicrophone & Reciprocity~\cite{hanneman2005introduction} & Balanced two-way exchanges (dialogue) & Predominantly one-way communication \\
\hline
\end{tabularx}
\end{threeparttable}
\end{table*}

\vspace{-0.3em}
\subsubsection{Network Measures for Behavioral Insights}

We compute four network metrics across the three modality-specific and fused sociograms: (1) eigenvector centrality~\cite{freeman1979centrality}, (2) clustering coefficient~\cite{watts1998collective}, (3) density~\cite{scott2011social}, and (4) reciprocity~\cite{hanneman2005introduction}. The first three apply to all four sociogram types, while reciprocity applies only to the directed conversation graph. A summary of how each metric maps onto interpretable aspects of group behavior is shown in \autoref{tab:network_metrics}.

To convert continuous metric values into interpretable categories, we apply a three-tiered scale using session-relative $z$-scores: \textbf{high} ($z \ge +1$, top 16\%), \textbf{medium} ($-1 < z < +1$, middle 68\%), and \textbf{low} ($z \le -1$, bottom 16\%), following the standard thresholds of the normal distribution. The $\pm 1$ threshold produced occupied bins across all 12 groups in our deployment, confirming that it captures meaningful variation rather than collapsing to a single category.

\vspace{-0.3em}
\subsubsection{Implementation Details}\label{sec:implementation_details}

For each window $k \in \{0, \dots, K\}$, we define an interval $[kS,\;kS+T]$ with window duration $T = 32$\,s and stride $S = 16$\,s. This 50\%overlap provides continuous temporal coverage with redundancy at window boundaries. Within each window we construct three$N\times N$ adjacency matrices $W^{(\mathrm{conv})},W^{(\mathrm{att})},W^{(\mathrm{prox})}$, where $N$ is the number of group members. The matrices are populated as follows, with threshold values discussed in \S\ref{sec:passive-sensing}.

\noindent\textbf{Conversation.}  For every detected speech segment $(s_{\mathrm{start}}, s_{\mathrm{end}}, p)$ attributed to participant $p$ that intersects the $\mathcal{I}_k$, we calculate the effective duration $\Delta$ as:
\begin{equation}
    \Delta = |[s_{\mathrm{start}}, s_{\mathrm{end}}] \cap \mathcal{I}_k|
\end{equation}

\begin{figure}
\centering
\includegraphics[width=\linewidth, trim={0cm 13cm 18.5cm 3cm},clip]{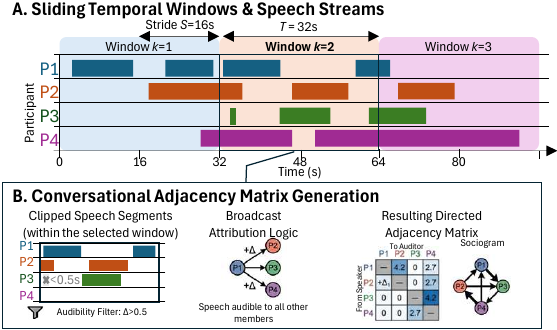}
\caption{Directed conversation sociogram construction. (A)~Speech segments are clipped to sliding windows ($T=\SI{32}{\second}$, $S=\SI{16}{\second}$) and filtered below \SI{0.5}{\second}. (B)~Broadcast attribution assigns each speaker's effective duration within the window as directed edges to all other members, yielding the weighted adjacency matrix visualized as a directed sociogram.}
\label{fig:convsocio}
\end{figure}

In our \SI{3.0}{\meter}$\times$\SI{1.5}{\meter} with $N=4$, we assume speech is audible to all other group members. If $\Delta \ge 0.5$\,s, we increment the directed edge weights from the speaker $p$ to all auditors $q$:
\begin{equation}
    W_{pq}^{(\mathrm{conv})} \leftarrow W_{pq}^{(\mathrm{conv})} + \Delta, \quad \forall q \in \{1, \dots, N\} \setminus \{p\}
\end{equation}
The resulting matrix $W^{(\mathrm{conv})}$ represents the cumulative directed speaking-listening duration within the window. A visual example of constructing the conversation sociogram is shown in \autoref{fig:convsocio}.

\noindent\textbf{Attention}. W we clip each user's gaze intervals to $\mathcal{I}_k$  and aggregate overlapping fixation durations between pairs $(i, j)$. Let  $\mathcal{O}{ij}^k$ denote the set of overlapping fixation intervals between users $i$
and $j$ within $Ik\mathcal{I}k$:
\begin{equation}
    W{ij}^{(\mathrm{att})} = W_{ji}^{(\mathrm{att})} = \sum_{\delta \in \mathcal{O}_{ij}^k} \delta, \quad \text{for } \delta \ge 13,\text{ms}
\end{equation}
This results in an undirected adjacency matrix where edges represent the total duration of joint attentional focus.

\noindent\textbf{Proximity}. We align headset poses at common timestamps within $\mathcal{I}_k$ an indicator function to count instances where the pairwise distance $d_{ij}$ falls within the proximity threshold:
\begin{equation}
    W_{ij}^{(\mathrm{prox})} = W_{ji}^{(\mathrm{prox})} = \sum_{t \in \mathcal{I}k} \mathbf{1}[d{ij}(t) \le 50,\text{cm}]
\end{equation}
This weight represents a sample count that reflects the frequency and duration of close-range physical interaction.

\noindent\textbf{Fusion.} To combine the three modality-specific matrices, we assemble an edge-weight matrix $X$ where rows correspond to directed dyad pairs and columns to modalities. For undirected modalities (attention and proximity), both directions receive identical weights. Each column is $z$-scored to equalize scale across modalities. We then fit PCA with a single component to the standardized matrix, square the resulting loadings to obtain variance-contribution proportions, and normalize these to sum to 1, yielding fusion weights $\alpha_{\mathrm{conv}}, \alpha_{\mathrm{att}}, \alpha_{\mathrm{prox}}$.  The fused sociogram is computed as a weighted linear combination of the standardized edge weights:

\begin{equation}
    W^{(\mathrm{fused})} = \sum_{m\in{\mathrm{conv},\mathrm{att},\mathrm{prox}}} \alpha_{m},Z^{(m)}
\end{equation}

where $Z^{(m)}$ denotes the $z$-scored edge weights for modality $m$. Squaring the loadings ensures all weights are non-negative. Because fusion operates on standardized values, the fused graph reflects relative interaction strength across modalities rather than raw duration counts.

\noindent\textbf{Runtime.} After filtering inactive windows, a typical group yields ${\sim}50$ structural windows (${\sim}14$ minutes of active interaction). Structural analysis of this volume completes in ${\sim}14$\,s on commodity hardware (Intel i7-14th gen, 32\,GB RAM).

\subsection{Temporal Analysis Module}\label{sec:temporalclustering}

Sociograms characterize interaction strength within each window but do not classify which behavioral pattern a given window represents or track how patterns recur across a session. In collaborative tasks, dyads do not maintain a single interaction style throughout a session but shift between qualitatively different modes on timescales of seconds. Our temporal analysis addresses this by segmenting each session's dyadic behavioral features into temporal phases via unsupervised clustering of time-series features. The complete pipeline is shown in the temporal block of \autoref{fig:sys-overview}.

\vspace{-0.3em}
\subsubsection{Feature Selection and Construction.}\label{sec:temporalfeature}

To capture the behavioral richness of dyadic interaction, we extract features spanning four dimensions: how pairs converse (turn-taking dynamics, speaking balance), how they move relative to each other (distance, approach, co-presence), what objects they jointly attend to and how broadly they explore task materials. We begin with a pool of 23 such features computed from moment-to-moment participant interactions at one-second intervals\footnote{The full set of candidate features: \emph{speak\_overlap}, \emph{speak\_only\_i}, \emph{speak\_only\_j}, \emph{speaker\_switch}, \emph{silence}, \emph{floor\_streak\_i}, \emph{floor\_streak\_j}, \emph{resp\_latency}, \emph{burst\_switch\_rate}, \emph{burst\_overlap\_rate}, \emph{dominance\_ratio}, \emph{entropy\_speaking}, \emph{bigram\_entropy}, \emph{fano\_switch}, \emph{dist\_mean}, \emph{prox\_binary}, \emph{approach\_rate}, \emph{dist\_accel}, \emph{dist\_jerk}, \emph{joint\_att\_cnt}, \emph{joint\_att\_dur}, \emph{shared\_att\_ratio}, \emph{material\_diversity}. Symmetric features (\_i/\_j pairs) are retained separately because dyad members are ordered, not interchangeable.}. We reduce this pool to a compact set through three successive filters that remove uninformative features, identify the most discriminative ones, and eliminate redundancy: (1)~a variance filter ($t=1\times10^{-5}$) removes near-constant features; (2)~K-Means clustering is fit on the data using the remaining features, with $k$ auto-selected via silhouette score, and a Random Forest classifier trained on the resulting cluster labels ranks features by importance, retaining the top 10; (3)~pairwise absolute correlations are computed among the retained features, and any feature with $r\geq 0.95$ is dropped. We adopt this unsupervised-then-supervised surrogate ranking because it surfaces features that best discriminate naturally occurring groupings in the data, without requiring predefined behavioral labels. This process yields 7 features.

This surrogate procedure serves only as a dimensionality reduction step; its cluster labels are discarded after feature ranking. The downstream temporal clustering that produces all reported results uses a separate convolutional-recurrent autoencoder (\S\ref{sec:sequence_encoding}) trained from scratch on only the 7 selected features, discovering its own latent space and cluster assignments. The 7 retained features span four behavioral dimensions:

\noindent\textbf{Verbal dynamics:} \emph{entropy\_speaking}, the evenness of turn-taking within the dyad (high = balanced exchange, low = silence or monopolization), and \emph{dominance\_ratio}, the imbalance in speaking time (0.5 = equal, deviations = asymmetric).

\noindent\textbf{Object diversity:} \emph{material\_diversity}, the count of distinct virtual objects both participants jointly attend to (high = broad surveying, low = sustained focus on few items).

\noindent\textbf{Proximity:} \emph{dist\_mean}, average dyadic distance (low = side-by-side, high = dispersed); \emph{prox\_binary}, co-presence within the defined proximity threshold (whether the pair is within arm's reach); and \emph{approach\_rate}, the rate of movement toward or away from one another (positive = convergence, negative = separation).

\noindent\textbf{Joint attention:} \emph{joint\_att\_cnt}, the number of joint gaze fixations on the same virtual object, capturing moments when both participants inspect the same item simultaneously.

All features are computed at one-second intervals, $z$-normalized per dyad, and aligned to a uniform temporal grid.

\vspace{-0.3em}
\subsubsection{Model Architecture.}\label{sec:sequence_encoding}
We treat each dyad's interaction over a segment as a $T \times F$ matrix, where $T$ is the number of \SI{1}{\second} windows per segment and $F = 7$ is the number of retained features. We select $T = 16$ and stride $S = 8$ (50\% overlap) via grid search, jointly optimizing silhouette score and reconstruction loss on held-out data (selection detailed in \S\ref{sec:window_stride}). The structural module operates at a different granularity ($T = 32$, $S = 16$); the two modules are not temporally aligned by design, as each optimizes its own window parameters independently. Each sequence is encoded by a convolutional-recurrent autoencoder~\cite{ma2021unsupervised, yin2020anomaly} into a 32-dimensional latent vector, which is then clustered via K-means with a deep embedded clustering objective ~\cite{xie2016unsupervised}.

\vspace{-0.3em}
\subsubsection{Window Length and Stride Selection.}\label{sec:window_stride}
For the temporal module (the structural module uses separate parameters, \S\ref{sec:structural}), we grid-searched window length $T \in {8, 16, 32}$ and stride $S \in {4, 8, 16}$ with the constraint $S \leq T/2$, yielding 6 valid combinations. Each configuration was evaluated in a coarse pass (1 epoch), which produced non-overlapping reconstruction loss ($0.47$--$0.66$) and silhouette score ($0.17$--$0.28$) ranges across configurations for stable ranking. The top 50\% were then refined (5 epochs), and candidates were ranked by an equally weighted combination of silhouette score and reconstruction loss. Stride was searched systematically across all window sizes, not only at the final $T$. Among the top-ranked configurations, $\langle T{=}16,\, S{=}8 \rangle$ achieved the best composite rank score. We adopt this configuration throughout, yielding $5{,}334$ dyadic windows across 12 groups (72 pairs).

\vspace{-0.3em}
\subsubsection{Implementation and Output.}
All computation is performed offline. To scale across many dyads or lengthy sessions, feature extraction, encoding, and clustering run in parallel. The full dataset ($5{,}334$ windows) was used without subsampling during final training; a fast evaluation mode that subsamples to $5{,}000$ windows is available for rapid iteration but was not used for reported results. On a consumer CPU (Intel i7-14th gen, 32\,GB RAM), the full pipeline processes 12 groups (6 dyads each, sessions averaging ${\sim}32$ minutes total) in ${\sim}80$ seconds total, with a per-group median of 3.25\,s ($\text{SD} = 0.33$, range $2.89$--$4.00$\,s). The bottleneck is $k$-selection, which accounts for ${\sim}99\%$ of per-group runtime. The module outputs a cluster label for each dyadic window, from which feature heatmaps and phase-aligned timelines can be derived. Cluster interpretations are reported in \S\ref{sec:eval-temporal}.

\section{Exploratory Deployment}

\subsection{Participants}
We recruited 48 participants (12 groups of 4; 36 male, 8 female; age mean ($\mu$) = 24.2, standard deviation (SD) = 4.7). 
Pre-study demographic questionnaires measured prior immersive-tech experience on 7-point Likert scales: MR ($\mu = 1.8, SD = 1.2$), AR ($\mu = 3.1, SD = 1.8$), VR ($\mu = 3.4, SD = 2.1$). We fixed the group size at four to maximize the number of dyadic interactions (six per group) while keeping computation tractable~\cite{university_82_nodate}.

\subsection{Materials}
We conducted the study in a \SI{3.0}{\meter}$\times$\SI{1.5}{\meter} space, cleared of materials to minimize distractions, where participants navigated and collaborated in close quarters. Each participant used a Meta Quest Pro headset~\cite{MetaQuestProa}, which captured and streamed eye gaze, audio, and 6DoF pose data over our local Wi-Fi network. We synchronized all devices with NTP (sub-\SI{100}{\milli\second} precision, as detailed in \S\ref{sec:passive-sensing}) and built the collaborative MR app in Unity with the Meta XR SDK~\cite{metaXRpackages}. All virtual objects were aligned to a shared coordinate frame so that each user viewed a consistent scene without additional prompts, cues, or enforced turn-taking.

\begin{figure*}[h]
\centering
\includegraphics[width=\linewidth]{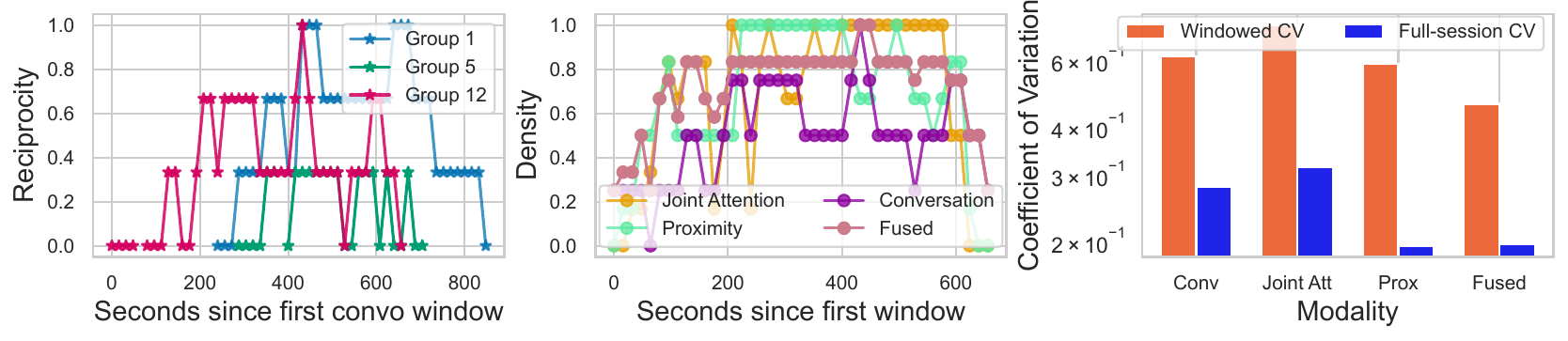}

\caption{Windowed-session analysis to assess behavioral patterns obscured by session-level aggregation. Conversation reciprocity for representative Groups $1, 5, 12$ (left), Group 12’s multimodal density trajectory (center), and density variation (right).}
\label{fig:windowed_variance}
\end{figure*}

\vspace{-0.3em}
\subsection{Collaborative Task}
We asked each group to sort the $28$ OASIS images~\cite{kurdi_introducing_2017} (prevalidated for pleasantness and arousal and free of graphic content) into six affective labels (angry, bored, relaxed, tense, pleased, frustrated) based on Russell's circumplex model~\cite{russell_circumplex_1980}. This image sorting task has been adopted in various group dynamics studies~\cite{gendron2014perceptions,chandio2025sensors}. 

The virtual images were scattered throughout, overlaid on the see-through physical space, with floating \emph{plates} labeled for each emotion hovering nearby. Without any time limit or scripted turns, participants freely approached and grabbed images, moved them to their chosen plates, and negotiated assignments through open-ended discussion. To sort an image, participants used a natural point-and-drag motion with their Meta Quest Pro controllers. They aimed at an image, held the grip button to \emph{pick it up}, guided it ok  the desired emotion plate, and released the button to lock it in place. Images only attach when positioned sufficiently close to a label, providing immediate visual confirmation. Only one person can manipulate a given image at a time, but different participants may simultaneously move other images within reach, mirroring the physical act of picking up and placing objects. This unstructured setting, where teams self-direct by clustering around images of interest, encourages natural decision-making, communication, and alignment as group members iteratively build consensus on each label~\cite{rorissaFreeSortingImages2004,bjerreCardSortingCollaborative2015,berlinFeedbackCooperationExperiment2024}. An egocentric view of the task is shown in \autoref{fig:sys-overview}. The average task completion time was 32.4 minutes (SD=8.4\text{SD} = 8.4
SD=8.4), but substantial portions of each session contained no detectable dyadic interaction (e.g., individual sorting, idle periods between subtasks). The pipeline retains only windows in which at least one modality registers nonzero dyadic activity, yielding a mean of 53 active structural windows per group (${\sim}15$ minutes of active interaction per session).

\vspace{-0.3em}
\subsection{Procedure}

Upon arrival, participants reviewed an IRB-approved information sheet and provided verbal consent, then completed a brief demographics survey. We handed out Meta Quest Pro headsets, guided each person through focus and fit calibration, and ran a short tutorial using two practice images and categories to teach the grab–drag–release interaction and category placement in MR. Next, groups tackled the main task, sorting $28$ images into $6$ emotion categories. We instructed them to \emph{work together to categorize these images by emotion, discuss, and reach agreement on each label}, with no time limit or performance feedback. Participants self-directed their collaboration, moving freely around the space and negotiating assignments until everyone confirmed consensus. 
The entire session, including setup, training, task, and wrap-up, took approximately $35 - 45$ minutes.

\vspace{-0.3em}
\subsection{Data Collection and Correspondence Checks}\label{sec:datacollection}
We captured multimodal data passively from each headset with post-hoc synchronization and ran it through \sysname's end-to-end pipeline (structural and temporal modules described in \S\ref{sec:structural} and \S\ref{sec:temporalclustering}, respectively ) to analyze group behavior. We assessed the pipeline's outputs through two complementary checks: behavioral \textbf{correspondence check} with observed activity and \textbf{internal consistency} (clustering stability and sociogram metric consistency across time and groups). 

For the \textbf{correspondence check}, one researcher reviewed time-aligned egocentric video from both members of each dyad. We sampled 100 windows (drawn from the middle portion of each session to avoid startup transients) and presented each alongside its predicted cluster label and characterization. Using synchronized video and audio from both participants' headsets, the coder judged whether the predicted behavioral pattern matched the observed interaction (match/mismatch/uncertain), and for mismatches recorded the observed cluster. This procedure yielded 83\% agreement between automated labels and human judgment; per-cluster precision, recall, and interpretation are reported in \S\ref{sec:eval-temporal}. Coding was performed by a single researcher due to data access restrictions under our IRB protocol.

We did not collect post-hoc subjective collaboration surveys. Post-hoc ratings of moment-level behavioral states are unreliable indicators of observed interaction dynamics~\cite{chandio2025reaction}, and our pipeline targets observable behavioral patterns rather than participants' internal experience. The correspondence check instead verifies that the pipeline's outputs align with behaviors visible on camera, without claiming access to a definitive ground truth for collaboration states.

\section{Results \& Analysis}\label{sec:results}
\subsection{RQ1: Multi-Granular Temporal Structural Patterns}
To answer \textbf{RQ1} presented in \S\ref{sec:intro}, we evaluate the structural module at two temporal granularities, examining sensing modality contributions to the fused sociogram through windowed metrics (\S\ref{sec:sessionvswindow}), cross-modal dependencies (\S\ref{sec:cross-modal}), and group-level network patterns (\S\ref{sec:netgroups}).

\vspace{-0.3em}
\subsubsection{Session-Level vs. Windowed Analysis}\label{sec:sessionvswindow}
We compared session-level and windowed interaction metrics presented in \autoref{tab:network_metrics}
across all groups. Session-level aggregation resulted in most graph metrics to  have near-identical values across groups ($CV:\ 0.000 - 0.095$), as participant pairs eventually interact over full sessions.
In contrast, \SI{32}{\second} sliding-window analysis (\S\ref{sec:implementation_details}) revealed substantial within-session variability ($CV:\ 0.268 - 0.895$), confirming that temporal resolution is necessary to capture meaningful fluctuations in collaborative structure (\autoref{fig:windowed_variance}, right panel).
Group~12's fused density trajectory over time illustrates this within-session variability (\autoref{fig:windowed_variance}, center panel) with per-modality density shifting significantly across windows that session-level aggregation obscures.
Per-group reciprocity and fused density values are reported in \autoref{tab:group_screening}

To assess conversation reciprocity variance, we analyzed three groups representing the observed spectrum: Group~5 (low; $\mu=0.179, \sigma=0.169$), Group~10 (moderate; $\mu=0.451, \sigma=0.273$), and Group~1 (high; $\mu=0.538, \sigma=0.292$). Across all groups, mean reciprocity ranged from $0.179$ to $0.538$ , offering granular differentiation that session-level aggregation would obscure. Their windowed reciprocity trajectories are illustrated in \autoref{fig:windowed_variance} (left panel).

\emph{Hence, this analysis reveals that session-level metrics compress critical temporal dynamics that windowed analysis successfully surfaces.}

\begin{table}[t]
\centering
\scriptsize
\caption{
Per-group conversation reciprocity and mean fused density computed from active-session windows only.
}
\vspace{-0.2cm}
\label{tab:group_screening}
\begin{tabular}{lccc}
\toprule
Group & Mean Reciprocity & $\sigma$ Reciprocity & Fused Density \\
\midrule
1 & 0.538 & 0.292 & 0.509 \\
2 & 0.305 & 0.234 & 0.424 \\
3 & 0.261 & 0.253 & 0.609 \\
4 & 0.400 & 0.216 & 0.669 \\
5 & 0.179 & 0.169 & 0.389 \\
6 & 0.313 & 0.281 & 0.621 \\
7 & 0.480 & 0.364 & 0.748 \\
8 & 0.308 & 0.233 & 0.716 \\
9 & 0.491 & 0.242 & 0.743 \\
10 & 0.451 & 0.273 & 0.723 \\
11 & 0.515 & 0.390 & 0.684 \\
12 & 0.350 & 0.268 & 0.720 \\
\bottomrule
\end{tabular}
\end{table}
\vspace{-0.3em}
\subsubsection{Modality Contributions and Cross-Modal Dependencies}\label{sec:cross-modal}
To evaluate the internal structure of the fused sociogram, we analyzed how individual sensing modalities contribute to the final representation.
\textbf{Principle Component Analysis (PCA) loadings and Variance.}
Across all groups, conversation and joint attention dominated the fusion with mean squared loadings of  $= 0.401$ and $0.409$, respectively. Proximity contributed least at $0.190$ The first principal component explained 57.2\% of the variance on average, indicating that the fused interaction density reflects integrated signals rather than a single dominant channel. This pattern is evident in Group 12's data (\autoref{fig:windowed_variance}, center panel), where fused interaction density incorporates multi-modal contributions rather than being driven by a single dominant channel.

\noindent\textbf{Cross-Modal Dependencies.} Analysis of $3{,}822$ dyad-windows showed high activity across modalities: conversation in 68.0\%, joint attention in 55.9\%, and proximity in 54.1\%. Proximity and attention exhibited the strongest coupling ($P(\text{prox} \mid \text{att}) = 0.70$; $P(\text{att} \mid \text{prox}) = 0.72$), while conversation conditioned on attention ($P(\text{att} \mid \text{conv}) = 0.67$) and proximity ($P(\text{prox} \mid \text{conv}) = 0.62$) was somewhat weaker. Lift analysis ($1.15-–1.29$) confirmed all pairs co-occur more often than chance, validating that these signals are physically and socially linked.

\noindent\textbf{Kernel PCA comparison.} To test for nonlinear structure, we compared linear PCA against Radial Basis Function (RBF) and polynomial kernels. Polynomial kernels yielded near-identical results (mean edge rank $\rho = 1.000$), and while RBF was less stable ($\rho = 0.905$) it preserved the top-ranked dyad in 11 of 12 groups. These results confirm the fused sociogram is robust to the fusion method, with linear PCA providing stable, interpretable weights. Structural differences under nonlinear fusion are discussed further in \S\ref{sec:discussion}.

\noindent\textbf{Leave-one-out ablation.} We evaluated modality redundancy by excluding one channel and measuring dyadic rank preservation. Removing any modality caused substantial disruption; dropping conversation resulted in the lowest preservation (mean $\rho = 0.605$; $\rho \geq 0.8$ in 6 of 12 groups), followed by proximity ($\rho = 0.567; \rho \geq 0.8$ in 4 of 12) and conversation ($\rho = 0.476; \rho \geq 0.8$ in 6 of 12). All three modalities showed mean $\rho$ below $0.7$, indicating that no single channel is redundant. We discuss the implications for modality selection in \S\ref{sec:discussion}.

\emph{From this analysis, we validate that the three sensing modalities provide complementary, non-redundant signals essential for capturing the full complexity of group structure.}

\vspace{-0.3em}
\subsubsection{Network Metrics Across Groups}
\label{sec:netgroups}
\textbf{Edge weight distributions.} Each modality produces a distinct profile. Proximity yields the highest total weight per group ($8592.6 \pm 4325.7$) and the widest range (max $2534.8 \pm 1773.7$). Joint attention has the lowest totals ($757.6 \pm 781.8$) but the highest edge inequality (Gini $= 0.272 \pm 0.122$). Conversation is the most uniform (Gini $= 0.235 \pm 0.076$), reflecting the broadcast edge attribution model described in \S\ref{sec:implementation_details}. Notably, the fused sociogram has the lowest Gini coefficient of all ($0.208 \pm 0.093$), indicating that integrating modalities produces a more balanced representation of group engagement than any single channel. All 12 fused graphs are complete (density $= 1.00$), meaning every participant pair has nonzero fused interaction strength.

\noindent\textbf{Structural role variation.} No single participant dominated across groups. Strength ratios between the most and least connected participants ranged from $1.10$ (Group 11) to $3.97$ (Group 8). 
Modality-level disagreement was significant: all three sensors agreed on the strongest dyad in only 2 of 12 groups, while in 5 groups, each modality identified a unique strongest pair. This divergence reinforces the necessity of the fused representation and cross-modal findings in \S\ref{sec:cross-modal}.

\noindent\textbf{Illustrative cases.} Contrasting patterns in Groups 8 and 11 highlight these dynamics. In Group 8, conversation was negatively correlated with both proximity ($\rho = -0.771$) and attention ($\rho = -0.714$), resulting in the lowest cross-modal agreement and the highest strength ratio. Conversely, Group 11 participation was nearly equal (ratio $1.10$), but the modalities still identified different strongest pairs, proving that modality disagreement persists even in egalitarian groups. 

\emph{Hence, we conclude that the fused representation provides a stabilized view of group structure that diverges from any single modality.}

\begin{figure}[t]
  \centering
  \includegraphics[width=\linewidth, trim={0 0 0 0},clip]{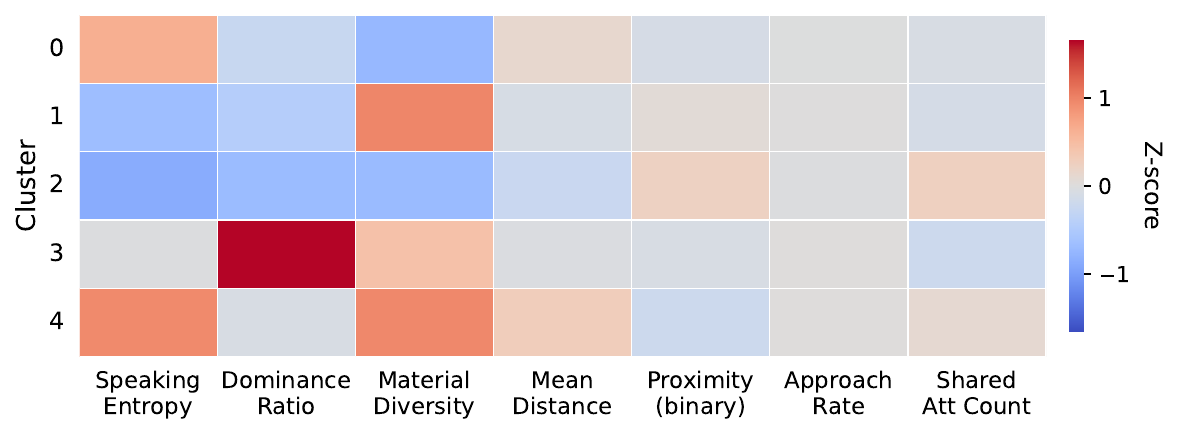}
  \caption{Heatmap of clusters (rows) vs. features (columns), where color intensity shows each feature’s deviation from its mean.}
  \label{fig:cluster_zheatmap}
\end{figure}

\subsection{RQ2: Emergent Behavioral Phases}\label{sec:eval-temporal}
To answer \textbf{RQ2}, we evaluate the behavioral patterns identified by the temporal clustering module. We discuss cluster selection (\S\ref{sec:clusterselection}), the interpretability of features via decision trees (\S\ref{sec:clusterinter}), and validation against egocentric video (\S\ref{sec:corrcheck}).

\vspace{-0.3em}
\subsubsection{Cluster Selection \& Characterization}
\label{sec:clusterselection}

The temporal clustering pipeline (\S\ref{sec:temporalclustering}) identified $k=5$ clusters with high stability (silhouette $= 0.81$, cross-seed $ARI = 1.0$) from $5{,}334$ dyadic windows (\S\ref{sec:window_stride})\footnote{The differing observation counts between sections reflect distinct windowing configurations optimized for the structural and temporal modules.}. 
To link clusters to behavioral patterns, we inspected their $z$-scored feature profiles (\autoref{fig:cluster_zheatmap}). 
Two of the seven selected features require interpretive context: entropy\_speaking and dominance\_ratio. High speaking entropy (${\sim}1.58$ bits) signifies balanced conversation with natural pauses, while low entropy (${\sim}0$) indicates state dominance, such as floor monopolization or silence. The dominance\_ratio measures speaking time symmetry: $0.5$ indicates balanced turns, with deviations categorized as near-balanced ($|z| < 0.3$), moderately asymmetric ($0.3 \leq |z| < 0.6$), or asymmetric ($|z| \geq 0.6$). The five clusters are characterized as follows:

\noindent \textbf{C0 (Balanced Narrow Focus, 28.1\%).} Characterized by balanced conversation (high speaking entropy, $z = +0.63$), but limited material engagement (low material diversity, $z = -0.73$) and near-balanced dominance ($z = -0.25$).

\noindent \textbf{C1 (Diverse Exploration, 15.9\%).} Pairs exploring a wide range of materials under a dominant conversational state with skewed speech. High material diversity ($z = +0.97$), low speaking entropy ($z = -0.66$), and moderately asymmetric dominance ($z = -0.45$).

\noindent \textbf{C2 (Concentrated Shared Focus, 23.4\%).} High joint attention and close physical proximity on a narrow set of materials, with asymmetric speech. Characterized by very low speaking entropy ($z = -0.85$), asymmetric dominance ($z = -0.69$), low material diversity ($z = -0.71$), and close proximity ($z = -0.24$).

\noindent \textbf{C3 (Guided Exploration, 18.8\%).} Strongly elevated dominance ratio ($z = +1.65$), one participant dominates the floor while both explore varied materials with moderately high material diversity ($z = +0.44$).

\noindent \textbf{C4 (Active Distributed Dialogue, 13.8\%).} Balanced, active conversation (high speaking entropy, $z = +0.95$) across diverse  materials ($z = +0.96$)), and balanced dominance with greater spatial separation ($z = +0.29$ for distance).

\vspace{-0.3em}
\subsubsection{Cluster Interpretability}
\label{sec:clusterinter}

We constructed a surrogate decision tree to distill each behavioral pattern into a rule hierarchy. Shapley additive explanations (SHAP) values~\cite{lundberg2017unified} confirmed that three features provide nearly all splitting power: material diversity (importance = 0.425), speaking entropy(0.320), and dominance ratio (0.251), with other features contributing less than 0.2\%. The root split utilizes dominance ratio to isolate Guided Exploration (C3). An edge case with extreme dist\_mean is reclassified as Active Distributed Dialogue (C4). For the remaining windows, speaking entropy distinguishes balanced from asymmetric speech, while material diversity separates narrow-focus from diverse-exploration patterns. This depth-4 tree reproduces cluster assignments with 98.9\% accuracy (macro $F1 = 0.987$). This high fidelity affirms that the unsupervised clusters map onto a transparent, logically consistent hierarchy of observable interaction behaviors.

\vspace{-0.3em}
\subsubsection{Distribution Across Dyads}
\begin{table}[t]
\centering
\scriptsize
\caption{Entropy of cluster membership across groups, pairs, and actors (lower values = more context-specific). Actor entropy is the mean of per-actor entropies for both dyad members.}
\label{tab:cluster_entropy}
\begin{tabular}{l>{\centering\arraybackslash}p{0.5cm}>{\centering\arraybackslash}p{0.5cm}>{\centering\arraybackslash}p{0.5cm}>{\centering\arraybackslash}p{0.53cm}}
\toprule
\textbf{Cluster} & \textbf{$N$} & \textbf{Group Ent.} & \textbf{Pair Ent.} & \textbf{Actor Ent.} \\
\midrule
C0 (Balanced Narrow Focus) & 1498 & 0.77 & 2.90 & 2.04 \\
C1 (Diverse Exploration) &  847 & 0.99 & 2.44 & 1.79 \\
C2 (Concentrated Shared Focus)          & 1250 & 0.97 & 2.83 & 2.17 \\
C3 (Guided Exploration)        & 1003 & 0.74 & 2.46 & 1.77 \\
C4 (Active Distributed Dialogue) &  736 & \textbf{0.63} & \textbf{2.13} & \textbf{1.47} \\
\bottomrule
\end{tabular}
\end{table}

To evaluate how interaction styles vary across social contexts, we computed the categorical entropy of cluster membership at the group, pair, and actor levels (\autoref{tab:cluster_entropy}). Lower entropy signifies behaviors concentrated in specific contexts, while higher scores indicate widely shared styles.

\noindent\textbf{Context-Specific Patterns:} Active Distributed Dialogue (C4) exhibited the lowest entropy across group (0.63), pair (2.13), and actor (1.47) levels, suggesting this balanced interaction style is highly dependent on specific team dynamics. Balanced Narrow Focus (C0) and Guided Exploration (C3) also showed low group entropy, indicating they are team-specific traits.

\noindent\textbf{Generalized Patterns:} Conversely, Diverse Exploration (C1) and Concentrated Shared Focus (C2) exhibited the highest group entropy ($0.99$ and $0.97$), demonstrating that these behaviors occur broadly across the participant pool regardless of group assignment.

\vspace{-0.3em}
\subsubsection{Behavioral Correspondence Check}\label{sec:corrcheck}

With the procedure described in \S\ref{sec:datacollection}, a single coder reviewed 100 stratified windows against synchronized egocentric video, achieving 83\% agreement with automated cluster assignments. Clusters 0 through 3 demonstrated high reliability, with F1 scores ranging from $0.83$ to $0.90$. Balanced Narrow Focus (C0) was the most accurately identified pattern with an F1 of $0.90$. Diverse Exploration (C1) and Concentrated Shared Focus (C2) both reached 0.95 recall. Active Distributed Dialogue (C4) achieved perfect precision but lower recall at $0.55$, primarily due to feature overlap with Guided Exploration. Overall macro-averaged metrics ($F1 = 0.83$) confirm that the temporal module reliably recovers dominant interaction patterns, with Active Distributed Dialogue presenting the most ambiguous boundary due to its shared features with neighboring clusters.

\subsection{RQ3: Complementarity of Structural and Temporal Analyses}

To answer \textbf{RQ3}, we examined whether structural and temporal modules capture overlapping or distinct aspects of collaboration.

\vspace{-0.3em}
\subsubsection{Cluster-Metric Cross-Tabulation}
To examine how temporal cluster assignments relate to structural network properties we cross-tabulated the five behavioral clusters against tertile-binned network metrics across 466 group-windows. Nine metrics showed significant associations with cluster membership at $p < 0.001$ (\autoref{tab:cluster_struct_assoc}), with joint attention eigenvector ($V = 0.33$) and fused reciprocity ($V = 0.32$) exhibiting the strongest effects. C2 was overrepresented in high joint-attention centrality (58.4\%), while C3 showed the highest fused reciprocity, C1 was almost absent from the high bin ($1.2\%$). For fused density, Balanced Narrow Focus (C0) and Guided Exploration leaned toward the low tertile ($51$--$53\%$), while C4 leaned high ($44.9\%$). 

Effect sizes remained small to moderate (Cramér's $V = 0.22$--$0.33$), indicating partial overlap rather than redundancy.

\begin{table}[t]
  \centering
  \scriptsize
  \caption{Structural metrics whose low/medium/high tertile distributions show significant association with behavioral clusters. Most metrics use the full aligned sample ($N{=}466$); reciprocity metrics have fewer observations because reciprocity is undefined in windows lacking directed edges. Of 14 candidate metrics, 5 collapsed to fewer than three tertile bins and were excluded; the remaining 9 all reached significance.}
  \label{tab:cluster_struct_assoc}
  \begin{tabular}{l>{\centering\arraybackslash}p{0.5cm}>{\centering\arraybackslash}p{0.5cm}>{\centering\arraybackslash}p{1cm}>{\centering\arraybackslash}p{1.3cm}}
    \toprule
    Metric (binned) & $N$ & $\chi^{2}$ & $p$ & Cram\'{e}r's $V$ \\
    \midrule
    Joint Attention Eigenvector & 466 &  99.40 & ${<}0.001$ & 0.33 \\
    Fused Reciprocity           & 341 &  68.12 & ${<}0.001$ & 0.32 \\
    Fused Density               & 466 &  76.82 & ${<}0.001$ & 0.29 \\
    Proximity Density           & 466 &  70.80 & ${<}0.001$ & 0.28 \\
    Conversation Density        & 466 &  65.08 & ${<}0.001$ & 0.26 \\
    Proximity Eigenvector       & 466 &  61.97 & ${<}0.001$ & 0.26 \\
    Fused Eigenvector           & 466 &  46.32 & ${<}0.001$ & 0.22 \\
    Conversation Eigenvector    & 466 &  45.70 & ${<}0.001$ & 0.22 \\
    Conversation Reciprocity\textsuperscript{\dag} & 316 & 30.24 & ${<}0.001$ & 0.22 \\
    \bottomrule
   \multicolumn{5}{l}{
  \parbox{0.8\linewidth}{
    \textsuperscript{\dag}\scriptsize Minimum expected cell count $= 2.8 < 5$; chi-squared approximation may be unreliable.
  }}
  \end{tabular}
\end{table}

\vspace{-0.3em}
\subsubsection{Structural Stability During Temporal Transitions}
\begin{figure}[t]
  \centering
  \includegraphics[width=\linewidth]{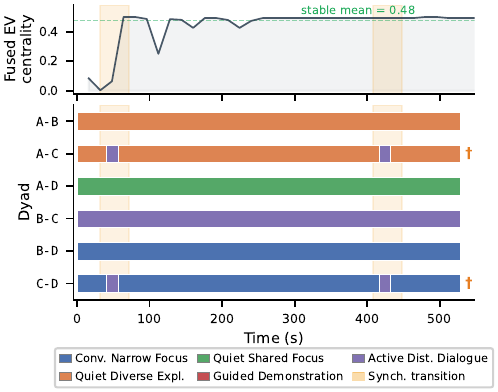}
  \caption{Fused eigenvector centrality (top) and temporal cluster assignments for all six dyads (bottom) in Group~12 over a \SI{528}{\second} session. Orange bands mark periods where two dyads (\dag) undergo synchronized transitions to \emph{Active Distributed Dialogue}; no structural metric registers a corresponding change ($0/14$, Mann--Whitney $U$, all $p > 0.07$ uncorrected).}
  \label{fig:timeline_case}
\end{figure}

To illustrate how the two modules complement each other, we present case analysis of Group 12 in \autoref{fig:timeline_case} with fused eigenvector centrality alongside temporal cluster labels for all six dyads in Group 12 over a \SI{528}{\second} session. 
Analysis shows that structural metrics remained stable while dyads concurrently occupied four distinct behavioral clusters with 100\% consistency: A–B in Diverse Exploration, A–D in Concentrated Shared Focus, B–C in Active Distributed Dialogue, and B–D in Balanced Narrow Focus. While structural analysis would characterize this state as uniformly cohesive, the temporal module identifies four distinct interaction patterns coexisting within it. Furthermore, synchronized transitions to Active Distributed Dialogue occurred without triggering significant shifts in any of the 14 structural metrics. Following an initial transient, fused eigenvector centrality stabilizes (mean $EV = 0.48$), masking these micro-level behavioral perturbations that only the temporal module detects.

This pattern generalizes to 8 of 12 groups, where three or more dyad-level clusters emerge under stable structural conditions. In Group 8, three dyads underwent synchronized transitions that structural metrics failed to register ($0/14$, $p > 0.20$ uncorrected). 

While cross-tabulation indicates partial overlap (Cramér's $V = 0.22$--$0.33$), these case studies confirm that temporal mode shifts often occur without detectable structural changes, proving the modules capture complementary aspects of group dynamics.

\section{Discussion and Limitations}\label{sec:discussion}
\subsection{Summary of Findings}
Three research questions guided this work. For RQ1, windowed structural analysis revealed within-session variability that session-level aggregation compresses entirely, and all three sensing modalities contribute meaningfully to the fused sociogram: conversation and attention co-dominate the PCA-derived fusion weights, while proximity, though the weakest contributor, is non-redundant, as removing any single channel disrupts dyadic rankings substantially. For RQ2, the temporal clustering module identified five behaviorally distinct interaction phases with perfect cross-seed stability, separable by just three features (material diversity, speaking entropy, and dominance ratio); a correspondence check against egocentric video confirmed 83\% agreement, indicating that the detected clusters map onto observable behavioral patterns. For RQ3, the two modules capture complementary rather than redundant dynamics: behavioral cluster transitions occur within structurally stable states, and while cross-tabulation reveals significant associations between temporal clusters and structural metrics, the effect sizes are moderate, indicating partial overlap rather than reducibility. Together, these findings suggest that structural and temporal views of collaboration each surface patterns the other misses, and that passive sensing from commodity headsets provides sufficient signal to support both levels of analysis.

\subsection{Multimodal Fusion and Modality Roles}
The PCA-based fusion assigns empirically derived weights that reflect each modality's contribution to shared variance, rather than assuming equal importance across channels. In our task, conversation and attention co-dominated the fusion weights while proximity contributed least, but this weighting is task-specific: physical assembly or navigation tasks, where spatial coordination is central, would likely shift the balance toward proximity. The kernel PCA comparison provides evidence that this linear fusion captures the dominant variance structure: the polynomial kernel recovered near-identical sociograms ($\rho > 0.999$), and even the less stable RBF kernel preserved the top-ranked dyad in 11 of 12 groups. However, fusion inherently compresses modality-specific information into a single summary view. Group~8 illustrates this cost: conversation was negatively correlated with both proximity and joint attention, a pattern invisible in the fused sociogram. Preserving per-modality views alongside the fused representation is therefore not optional but necessary for analysts investigating how different behavioral channels relate to one another. The pipeline supports this by computing and storing all four network types independently, with fusion as an additional layer rather than a replacement.

\vspace{-0.3em}
\subsection{Validation Without Ground Truth}
Moment-to-moment collaboration states are not objectively observable constructs: they are shaped by internal states, social context, and subjective interpretation~\cite{o2020intercoder}, and even trained raters typically achieve only moderate agreement when coding behavioral observations from video~\cite{hallgren2012computing}. Post-hoc self-reports, while valuable for capturing participants' subjective experience, are unreliable indicators of the micro-level behavioral dynamics that automated sensing detects~\cite{chandio2025reaction}. This is not a limitation specific to \sysname\ but a domain-inherent constraint shared across social signal processing and collaboration analytics. Prior work in these fields validates automated behavioral analysis through internal consistency, temporal stability, and expert correspondence rather than per-frame ground truth: Pentland's sociometric badge research evaluated sensing outputs against organizational outcomes and observer ratings rather than moment-level labels~\cite{pentland2012new}, and Echeverria et al.'s collaboration translucence framework relied on expert interpretation of system outputs to establish face validity~\cite{echeverria2019towards}. We position \sysname\ in this tradition as an exploratory pipeline that assists analysts rather than a classifier requiring labeled ground truth. The relevant validation criteria are reliability (cross-seed ARI $= 1.0$, surrogate tree fidelity $= 98.9\%$) and behavioral correspondence (83\% agreement with observed video), not correctness against subjective labels. We are candid about the limits of this correspondence check: a single coder reviewed 100 of 5,334 windows under IRB access restrictions, with no inter-rater reliability data. This establishes behavioral plausibility, not statistical generalization. Expanding coverage, adding a second coder, and computing inter-rater agreement are necessary steps before the cluster labels can be treated as validated categories.

\vspace{-0.3em}
\subsection{Limitations}
The results reported here are specific to a co-located, object-sorting task performed by four-person groups. The five behavioral clusters, relative modality weightings, and cross-modal dependency patterns should not be assumed to transfer to tasks with different interaction demands, such as debates, remote collaboration, or physical assembly. The seven retained temporal features were empirically selected from this dataset; tasks with scripted turns, asymmetric roles, or different physical constraints may require different feature sets. With 12 groups (48 participants), the sample is sufficient for pipeline demonstration and internal validation but limits statistical generalization, as reflected in the small-to-moderate cross-tabulation effect sizes.
The three sensing modalities (gaze, audio, position) were chosen deliberately for lightweight deployment on commodity headsets, but they omit hand gestures, facial expressions, and virtual object manipulation context that shape MR collaboration. The modular architecture supports extension with additional channels, but current results reflect only these three. Conversation edge attribution broadcasts each utterance to all group members, which may overestimate conversational engagement in subgroup-specific dialogue. On the validation side, the correspondence check relied on a single coder reviewing 100 of 5,334 windows with no inter-rater reliability data. Active Distributed Dialogue (C4) showed the lowest recall ($0.55$, $F1 = 0.71$), with most misclassified windows reassigned to neighboring clusters, indicating that the boundary between balanced high-entropy conversation and adjacent patterns needs refinement. These constraints define the scope within which the pipeline's outputs should be interpreted and point toward the extensions needed before deployment in applied settings.

\section{Conclusion}
We presented \sysname, a passive sensing pipeline that derives both structural and temporal views of group collaboration from commodity MR headset data without external instrumentation or manual annotation. By integrating automated multimodal sociograms with unsupervised deep clustering, the system reveals within-session variability and distinct behavioral phases that traditional session-level aggregation obscures. Our findings establish that headset embedded sensors can provide sufficient signal to identify complex group organization and social roles.

The identified five-phase behavioral taxonomy and modality weightings are context-specific to co-located, four-person sorting tasks. While cross-tabulation demonstrates the modules provide complementary, non-reducible insights, further validation across diverse collaborative settings and richer sensing modalities is required. Ultimately, \sysname provides a foundation for scalable, automated tools that augment analyst interpretation, advancing the state-of-the-art in immersive collaboration research.

\acknowledgments{
This work is supported by the U.S. National Science Foundation (NSF) under grant number 2339266 and 2237485.}

\bibliographystyle{abbrv-doi}
\bibliography{paper}

\end{document}